\documentclass[preprint,12pt]{elsarticle}
\newcommand{\ie}{\emph{i.e.}, }
\newtheorem{tw}{Theorem}[section]

\newtheorem{ass}{Assumption}[section] 

\usepackage{amssymb}

\journal{Statistics and Probability Letters}

\begin{document}

\begin{frontmatter}



\title{On Geometric Ergodicity of Skewed - SVCHARME models}


\author[agh]{Jerzy P. Rydlewski\fnref{ry}}
\author[uj,uek]{Ma\l{}gorzata Snarska\fnref{sn}}
\address[agh]{AGH University of Science and Technology, Faculty of Applied Mathematics, A. Mickiewicza 30, 30--059 Krak\'{o}w,
         Poland}
\address[uj]{Marian Smoluchowski Institute of Physics and Mark Kac Complex Systems Research Centre, Jagiellonian University, Reymonta 4, 30--059 Krak\'{o}w,
             Poland}
\address[uek]{Cracow University of Economics, Chair of Econometrics and Operations Research, Rakowicka 27, 31--510 Krak\'{o}w, Poland}

\fntext[ry]{ry@agh.edu.pl}
\fntext[sn]{snarskam@uek.krakow.pl}

\begin{abstract}
Markov Chain Monte Carlo is repeatedly used to analyze the properties of intractable distributions in a convenient way.
In this paper we derive conditions for geometric ergodicity of a general class of nonparametric stochastic volatility models with
skewness driven by hidden Markov Chain with switching.\\
\end{abstract}
\begin{keyword}

Markov switching\sep geometric ergodicity\sep irreducibility\sep mixture models\sep asymmetric stochastic volatility

\end{keyword}

\end{frontmatter}

\section{Introduction}
\label{intro}

The asymmetry in the intertemporal relation between volatility and the stock return is the purpose of active study. In general, asymmetric effects in
volatility denote, that the effects of positive returns on volatility are different from those of negative returns of a similar magnitude.
Asymmetry is also sometimes referred to us as a negative relation between the ex-post volatility in the rate of returns on equity
and the current value of the equity,\ie the commonly investigated leverage effect. It is therefore necessary to stress, that although leverage effect
exhibits asymmetry, not all asymmetric effects display leverage.

Two different approaches are of considerable interest, namely the conditional and stochastic volatility models.
In the class of conditional volatility, ARCH specifications that have been developed to capture asymmetric effects are: the Exponential
GARCH (EGARCH) model of Nelson (1991) \cite{Nel91} and the GJR model of Glosten, Jagannathan and Runkle (1993) \cite{GloJagRun93}.
In the manner depicted earlier, EGARCH model can describe leverage whereas the GJR model can capture asymmetric effects but not leverage.
On the other hand, stochastic volatility (SV) models have become a natural alternative to time-varying volatility of the ARCH family.
The asymmetric property within the stochastic volatility framework is established on a straightforward correlation between the innovations in both returns
and volatility. Unlike ARCH type models, the SV models allow volatility dynamics to follow some latent stochastic
processes based on volatility's unobservable nature.
Basic stochastic volatility process (BSV) was originally developed by Hull and White (1987) \cite{HulWhi87} and Jacquier et al. (2004)\cite{Jac04}
in continuous time framework as a mode to capture negative correlation between the innovation terms in the Black-Scholes option pricing formula.
This basic version of the SV model
assumes stationary error disturbances and uncorrelated Gaussian white noise processes (Harvey and Ruiz, 1993 \cite{HarRui93}).
In empirical research, extensions of a simple discrete time model due to Taylor (1986)\cite{Tay86} have been analyzed by Wiggins (1987)\cite{Wig87} and
Harvey and Shephard (1996)\cite{She96} in order to accommodate the direct correlation. This generalization, based on the immediate correlation between
the innovation is known as the SV with leverage (SV-L) model (Asai and McAleer, 2006 \cite{AsaMcA06}).
Other asymmetric models were suggested by Danielsson (1994)\cite{Dan94}, which was similar in the spirit to that of the EGARCH model. Nelson (1991)\cite{Nel91}
used the absolute value function to capture the sign and magnitude of the previous value of normalized returns
in accommodating asymmetric behaviour into an ARCH-type model. Danielsson (1994)\cite{Dan94} used the absolute value function as in Nelson (1991) \cite{Nel91},
but incorporated the observed return into the SV specification as it is not computationally straightforward in the SV framework to incorporate the
normalized disturbances, in a SV with leverage and size effects (SV-LSE) model.
So, Li and Lam (2002) \cite{SoLiLam02} considered a different type of threshold effects model in which the breaks in the constant and autoregressive
parameter in the SV
equation depend on the signs of the previous returns. An alternative form of asymmetry can be based on threshold effects, as proposed in
 Glosten, Jagannathan and Runkle (1992)\cite{GloJagRun93}
in the context of conditional volatility models. A variety of symmetric and asymmetric, univariate and multivariate, conditional and stochastic volatility
models is analyzed in McAleer (2005)\cite{McA05}.
Excess kurtosis and skewness in the residuals of the ''basic'' SV model has triggered the use of either the mixtures of normals or the central t--distribution
(Geweke, 1994 \cite{Gew94};
Shephard and Pitt, 1997 \cite{ShePit97}). Hansen (1994)\cite{Han94} has considered skewness in a GARCH model using skew t-distribution errors allowing for both
skewness and heavy
tails to co-exist in a time-varying volatility specification. Also, Cappuccio et al. (2004)\cite{Cap04} have managed to generate skewness and excess kurtosis in
the
conditional distribution of returns by assuming a skew-GED distribution. Volatility asymmetry, initially interpreted as the leverage effect, is expressed by
the
correlation of the returns with the future volatility. Under this assumption, when mean returns are negatively correlated with contemporaneous volatility,
the negative (positive) return shocks are associated
with increased (decreased) future volatility. Harvey and Shephard (1996)\cite{She96} have developed the asymmetric SV (ASV-t) model,
 which guarantees the existence of the martingale property. Recently, Wang et al. (2011)\cite{Wan11} have developed a heavy-tailed SV
model with leverage effect, where a bivariate t-distribution, in the form of a scale mixture of normal, is used to model the
error innovations of the return and volatility equations.
Tsiotas (2012)\cite{Tsio12} introduced generalised ASV models that incorporate financial data's stylized facts including excess
kurtosis, skewness, and volatility asymmetry, using three alternative distributions: the noncentral-t
(NCT), the skew-normal (SN), and the skew-t (ST) distributions. These  ASV models  incorporate both the Gaussian and the non-Gaussian assumption in the same
specification, while
taking into account the leverage effect. The asymmetric central-t distributed SV (hereafter ASV-t) model
nests in the asymmetric noncentral-t distributed SV (hereafter ASV-nct) model. The other one is the asymmetric skewnormal
SV (hereafter ASV-sn) model. Finally, in the asymmetric skew-t SV (hereafter ASV-st) model both
excess kurtosis and skewness co-exist.
The first multivariate SV model was proposed by Harvey, Ruiz and Shephard (1994)\cite{HarRuiShe94}, who specified the model in terms of instantaneous
correlations in the
mean and volatility equations. However, their estimation technique was based on the inefficient quasi-maximum likelihood (QML) procedure.
Danielsson (1998)\cite{Dan98}
suggested a multivariate SV-L model based on the specification considered by Harvey, Ruiz and Shephard (1994)\cite{HarRuiShe94}, but only estimated a
symmetric version of the model. Shephard (1996) \cite{She96} proposed a one factor multivariate SV model.

Parametric and log-volatility estimation of the asymmetric stochastic volatility models is usually implemented in a Markov Chain Monte Carlo (MCMC) set-up.
By overcoming the problem of likelihood non-existence in a close form solution, because of normality departures in the observed process, the
Metropolis-Hastings algorithm is common in the MCMC egine. These algorithms provide a reliable structure to explore the intractable distributions.
In any MCMC analysis, the convergence rate of the associated Markov Chain is of practical and theoretical importance.
Geometric ergodicity is essential in convergence of Markov Chains to their stationary distributions.
A geometrically ergodic chain converges to its target distribution at a geometric rate. In addition to ensuring the rapid convergence required for useful
simulation, geometric ergodicity is a key sufficient condition for the existence of Central Limit Theorems and consistent estimators of Monte Carlo standard
errors, \ie geometric ergodicity justifies the applicability of the Central Limit Theorems to ergodic averages along the path of the chain.
The main aim of this paper is to provide a formal proof of geometric ergodicity for general nonparametric model, that incorporates both
the asymmetry and is the natural extension to CHARME models provided by Stockis et al. 2010.
\section{The Skewed Stochastic Volatility-CHARME model}
Consider first a nonlinear $p$th order autoregressive process defined by
\begin{equation}
X_t=m(X_{t-1},...,X_{t-p})+\sigma(X_{t-1},...,X_{t-p})\epsilon_t+A(X_{t-1},...,X_{t-p})\iota_t^2
\label{eq:1},
\end{equation}
where $m$, $\sigma$ and $A$ are unknown, real-valued Borel measurable functions. The $\epsilon_t$ and $\iota_t$ are mutually independent identically distributed
zero-mean innovations.
Since, it is unremarkably not realistic to postulate that the observed process has the same trend function $m$, volatility function $\sigma$ or skewness
function $A$ at each time instant,
we consider a class of nonparametric time-series models, where between random change points, the process we observe is piecewise stationary. The dynamics of
$\{X_t\}$
is driven by a hidden Markov chain $\{Q_t\}$ with values in a finite set $\{1,2,...,K\}$. Our model is defined as follows:

\begin{equation}
X_t=\sum_{k=1}^K S_{tk}\left(m_k(X_{t-1},...,X_{t-p})+\sigma_k(X_{t-1},...,X_{t-p})\epsilon_t+A_k(X_{t-1},...,X_{t-p})\iota_t^2\right)
\label{eq:2},
\end{equation}

where

$$ S_{tk}=\left\{ \begin{array}{cccccc}\ 1,  \textit{ for } Q_t=k                                                                                                                                         \\                                                                                                                                                                 0, \textit{ otherwise }                                                                                                                                           \end{array}                                                                                                                                                       \right.$$

$m_k$, $\sigma_k$ and $A_k$, $k=1,2,...,K$ are unknown functions, $\epsilon_t$ and $\iota_t$ are independent and identically distributed random variables with
mean zero and they are mutually independent. \subsection{Model definition}
We make the following assumptions, adopting Stockis et al. \cite{StoFra} point of view:
\begin{ass}
 The process $\{Q_t\}$ is a first - order strictly stationary Markov chain which is irreducible and aperiodic with probability distribution
 $\left(\pi_1,\pi_2,...,\pi_K \right)$ and transition
probability matrix $A=\{a_{ij}\}_{1\leq i,j\leq K}$.
\end{ass}
\begin{ass}
Let $\mathbb{G}_{t-1}=\sigma\{X_s,s\leq t-1\}$ be the $\sigma$-algebra generated by $\{X_s,s\leq t-1\}$ and let $G_{t-1}\in \mathbb{G}_{t-1}$. Then $P(Q_t=j|
Q_{t-1}=i, G_{t-1})=P(Q_t =j| Q_{t-1}=i)$ for all $i,j.$ In other words, the hidden process $Q_t$ is independent of the past observations of $\{X_{t}\}$ given
its own past.

\end{ass}
\begin{ass}
 $Q_t$ is uncorrelated with the $\epsilon_t$ and $\iota_t$, given $(Q_{t-1},X_{t-1},X_{t-2},...)$.
\end{ass}
\begin{ass}
 Both $\epsilon_t$ and $\iota_t$ are independent of $X_{t-1},X_{t-2},...$.
\end{ass}
\begin{ass}
 The functions $m_k$, $\sigma_k$ and $A_k$ are bounded on compact sets for all $k$.
\end{ass}
\begin{ass}
 The iid random variables $\epsilon_t$ have a continuous and everywhere positive density $f$ and the iid random variables $\iota_t$ have a continuous and
 everywhere positive density $g$.
\end{ass}
Without the loss of generality, we restrict ourselves to the case $p=1$, \ie $m_k$, $\sigma_k$ and $A_k$ are functions on the real line. We assume that
$A_k(x)>0.$
We should make the following assumptions
\begin{ass}
 The iid random variables $\epsilon_t$ and $\iota_t$ have mean zero and variance equal to $1$. Moreover, there exists the fourth moment $E(\iota_t^4)=\kappa$.
\end{ass}
\begin{ass}
$$
\max_{i\in\{1,2,...,K\}}\lim_{|x|\to+\infty}\sup \frac {\sum_{k=1}^{K}a_{ik}\left( m_k^2(x)+\sigma_k^2(x)+\kappa A_k^2(x)+2m_k(x)A_k(x)\right)}{x^2}<1
$$
\end{ass}
Let $S_t=(S_{t1},S_{t2},...,S_{tK})^T$.
We conclude that, under assumptions 1 -- 4, the process $$Z_t=(S_t,X_t)^T,$$
which represents the transformed mixture process, is a Markov chain as well.
\begin{tw}
Under Assumptions 1 -- 8, the process $\{Z_t\}$ is geometrically ergodic.
\label{tw1}
\end{tw}
Proof. We shall prove that the conditions of Theorem 15.0.1 (iii) of Meyn and Tweedie \cite{MeyTwe93} are satisfied.
\newline 1. The process  $Z_t$ is $\varphi$-irreducible if we take $\varphi$ as a product of the stationary probability distribution measure of $\{Q_t\}$ on
$\{1,2,...,K\}$
and the Lebesgue measure on $\mathbb{R}$. Let $A=A_1\times A_2$ be such that $\varphi(A)>0$. $A_1$ contains at least one integer k between 1 and K.
 It suffices to prove that there exists t such that for all k and l
$$P\left((S_{t+1}, X_{t+1})\in \{e_k\}\times A_2 \vert S_1=e_l, X_1=x\right)>0$$
with $e_k$ denoting a unit vector with the $k$th component equal to 1.
\newline We obtain
$$P\left((S_{2}, X_{2})\in \{e_k\}\times A_2 \vert S_1=e_l, X_1=x\right)=$$
$$P\left(Q_{2}=k, X_{2}\in A_2 \vert S_1=e_l, X_1=x\right)=$$
$$P\left( X_{2}\in A_2 \vert Q_2=k, S_1=e_l, X_1=x\right)P\left(Q_{2}=k \vert Q_1=l, X_1=x\right)=$$
$$a_{lk}P\left(m_k(x)+\sigma_k(x)\epsilon_2+A_k(x)\iota_2^2\in A_2\right)=$$

$$a_{lk}\int_{A_2}\int_{\mathbb{R}}\frac{1}{\sigma_k(x)}f\left(\frac{u-y-m_k(x)}{\sigma_k(x)}\right)
\frac{g\left(\sqrt{\frac {y}{A_k(x)}}\right)+g\left(-\sqrt{\frac {y}{A_k(x)}}\right)}{2\left(\sqrt{\frac {y}{A_k(x)}}\right)}\mathbb{I}_{\{y>0\}} dydu $$
$$=a_{lk}h_k(x),$$
where $h_k(x)>0.$

The one but last equality we obtain, because
the density of $m_k(x)+\sigma_k(x)\epsilon_2$ is $$\overline{f}(y)=\frac{1}{\sigma_k(x)}f\left(\frac{y-m_k(x)}{\sigma_k(x)}\right) $$
and the density of $A_k(x)\iota_2^2$ is
$$\overline{g}(y)= \frac{g\left(\sqrt{\frac {y}{A_k(x)}}\right)+g\left(-\sqrt{\frac {y}{A_k(x)}}\right)}{2\left(\sqrt{\frac
{y}{A_k(x)}}\right)}\mathbb{I}_{\{y>0\}}.$$
Then we use a well--known formula for convolution of the distributions, \ie $$\left(\overline{f}\ast
\overline{g}\right)(y)=\int_{\mathbb{R}}\overline{f}(u-y)\overline{g}(y)du.$$

Similarly,
$$P\left((S_{3}, X_{3})\in \{e_k\}\times A_2 \vert S_1=e_l, X_1=x\right)=$$
$$P\left(Q_{3}=k, X_{3}\in A_2 \vert S_1=e_l, X_1=x\right)=$$
$$P\left( X_{3}\in A_2 \vert Q_3=k, Q_1=l, X_1=x\right)P\left(Q_{3}=k \vert Q_1=l, X_1=x\right)=$$
$$\sum_{j=1}^Ka_{lj}a_{jk}\int_{A_2}\int_{\mathbb{R}}P(x,dy|Q_3=k,Q_1=j)P(y,du|Q_3=k,Q_1=j)=$$
$$\sum_{j=1}^Ka_{lj}a_{jk}h_{jk}(x)$$

where

$$h_{jk}(x)=\int_{A_2}\int_{\mathbb{R}}\int_{\mathbb{R}}\int_{\mathbb{R}}
\frac{1}{\sigma_k(y)}f\left(\frac{u-v-m_k(y)}{\sigma_k(y)}\right)
\frac{g\left(\sqrt{\frac {v}{A_k(y)}}\right)+g\left(-\sqrt{\frac {v}{A_k(y)}}\right)}{2\left(\sqrt{\frac {v}{A_k(y)}}\right)}\mathbb{I}_{\{v>0\}}
\cdot $$
$$\cdot \frac{1}{\sigma_j(x)}f\left(\frac{y-w-m_j(x)}{\sigma_j(x)}\right)
\frac{g\left(\sqrt{\frac {w}{A_j(x)}}\right)+g\left(-\sqrt{\frac {w}{A_j(x)}}\right)}{2\left(\sqrt{\frac {w}{A_j(x)}}\right)}\mathbb{I}_{\{w>0\}} dwdvdydu
$$

and $h_{jk}(x)>0.$
\newline Finally, we obtain
$$P\left((S_{t+1}, X_{t+1})\in \{e_k\}\times A_2 \vert S_1=e_l, X_1=x\right)=\sum_{j_1,...j_{t-1}}^Ka_{lj_1}...a_{j_{t-1}k}h_{j_1,...,j_{t-1}}(x)$$
which is strictly positive for some t because of the irreducibility of $\{Q_t\}$ and the fact that $h_{j_1,...,j_{t-1}}(x)>0$.
\newline 2. Analogously can be proven the aperiodicity of $\{Z_t\}$.
\newline 3.
Irreducibility and aperiodicity ensure that every petite set is small.
It can be shown that in the model we consider, each compact set is small.
It suffices to prove that for every compact set $B$ such that $\varphi(B)>0$ and for every bounded Borel set $A=A_1\times A_2$ with $\varphi(A)>0$ there exists
$t$
such that
$$\inf_{x\in B} P\left((S_{t+1}, X_{t+1})\in \{e_k\}\times A_2 \vert S_1=e_l, X_1=x\right)>0.$$
Proceeding as in the proof of irreducibility, we need to show that
$$\inf_{x\in B} \sum_{j_1,...j_{t-1}}^Ka_{lj_1}...a_{j_{t-1}k}h_{j_1,...,j_{t-1}}(x)>0$$
for some $t$.
Continuity of $h_{j_1,...,j_{t-1}}(x)$ and irreducibility of $\{Q_t\}$ ensure us that the last infimum is positive (see \cite{BhaLee95, BhaLee99}).
\newline 4. We will apply drift criterion of Theorem 15.0.1 (iii). We need to show that there exist $L>0$, $\beta>0$ and a function $V(Z)>1$ such that for
$\|Z_{t-1}\|>L$ we have
$$\frac{E\left(V(Z_t) \left|\right. Z_{t-1}=(e_l, x) \right)-V\left(e_l, x\right)}{V\left(e_l, x\right)}\leq -\beta.$$
Let $V(Z_t)=1+X_t^2.$
We obtain
$$\frac {E\left(V(Z_t) \left|\right. Z_{t-1}=(e_l, x) \right)-V\left(e_l, x\right)}{V\left(e_l, x\right)}=$$
$$\frac{\sum_{k=1}^K \left( m_k^2(x)+\sigma_k^2(x)+\kappa A_k^2(x)+2m_k(x)A_k(x)\right)E\left(S_{tk} \left|\right. S_{t-1}=e_l\right)-x^2}{1+x^2}\leq $$
$$\frac{\sum_{k=1}^K \left( m_k^2(x)+\sigma_k^2(x)+\kappa A_k^2(x)+2m_k(x)A_k(x)\right)a_{lk}}{x^2}-1$$
The conclusion is obtained by Assumption 8.
$\hfill \square$ \\

\section{Summary and remarks} We have derived sufficient conditions for geometric ergodicity of a general class of asymmetric nonparametric stochastic processes
with stochastic volatility.
It is natural to ask, whether other kinds of ergodicity (eg. polynomial ergodicity) can also be related to the skewed SV-CHARME model. It is the purpose of
our future work.
\section{Acknowledgements} MS acknowledges the support of EFS Human Capital grant (POKL.08.02.01-12-073/10-00/41-dz./2012).

\bibliographystyle{elsarticle-harv}

\end{document}